\documentclass[%
reprint,
showkeys,
nofootinbib,
amsmath,amssymb,
aps,
prl,
]{revtex4-1}
\usepackage[marginal]{footmisc}
\usepackage{graphicx}
\usepackage{dcolumn}
\usepackage{bm}
\usepackage[colorlinks,allcolors=black,CJKbookmarks=True]{hyperref}

\begin{document}
	
	\title{Constraining first-order phase transitions with curvature perturbations}
	
	\author{Jing Liu$^{1,2}$}
	\email{liujing@ucas.ac.cn}
	
	\author{Ligong Bian$^{3,4}$}
	\email{lgbycl@cqu.edu.cn}
	
	\author{Rong-Gen Cai$^{5,2,1}$}
	\email{cairg@itp.ac.cn}
	
	\author{Zong-Kuan Guo$^{5,2,1}$}
	\email{guozk@itp.ac.cn}
	
	\author{Shao-Jiang Wang$^{5}$}
	\email{schwang@itp.ac.cn}
	
	\affiliation{$^{1}$School of Fundamental Physics and Mathematical Sciences, Hangzhou Institute for Advanced Study, University of Chinese Academy of Sciences, Hangzhou 310024, China
	}
	
	\affiliation{$^{2}$School of Physical Sciences, University of Chinese Academy of Sciences,
		No.19A Yuquan Road, Beijing 100049, China}
	
	\affiliation{$^{3}$Department of Physics, Chongqing University, Chongqing 401331, China}
	
	\affiliation{$^{4}$Center for High Energy Physics, Peking University, Beijing 100871, China}
	
	\affiliation{$^{5}$CAS Key Laboratory of Theoretical Physics, Institute of Theoretical Physics,
		Chinese Academy of Sciences, P.O. Box 2735, Beijing 100190, China}
	
	\begin{abstract}
		We investigate the  curvature perturbations induced by the randomness of the quantum tunneling process during cosmological first-order phase transitions~(PTs) and for the first time ultilize curvature perturbations 
		to constrain the PT parameters. 
		We find that the observations of the cosmic microwave background spectrum distortion and the ultracompact minihalo abundance can give strict constraints on the PTs below $100$GeV, especially for the 
		low-scale PTs and the weak PTs. The current constraint on the PT parameters is largely extended by the results in this work.
	\end{abstract}
	
	\maketitle
	\emph{Introduction}. 
	The cosmological first-order phase transitions~(PTs) are expected to take place in many well-motivated new physics models~\cite{Losada:1996ju,Cline:1996mga,Laine:1996ms,Bodeker:1996pc}. During a first-order PT, at least two non-degenerated local minima appear  in the effective potential simultaneously, and the metastable phase decays due to quantum tunneling or thermal fluctuations~\cite{Coleman:1977py,Callan:1977pt,Linde:1980tt,Linde:1981zj}. True vacuum bubbles copiously nucleate and then expand until they collide with each other, releasing the vacuum energy into bubble walls and background plasma. As a violent process in the early Universe, the PTs are expected to produce observable relics including gravitational waves~(GWs)~\cite{Caprini:2015zlo,Caprini:2019egz}, primordial magnetic fields~\cite{Vachaspati:1991nm,Di:2020ivg} and baryon asymmetry~\cite{Morrissey:2012db}. Since the electroweak PT in the Standard model of particle physics is crossover~\cite{DOnofrio:2014rug}, the experiments aiming at observing the relics of the first-order PTs help to determine or constrain the parameters of new physics models. Observing gravitational waves produced during the PTs is one of the main scientific goals of various observational projects, such as LISA~\cite{Audley:2017drz}/$Taiji$~\cite{Guo:2018npi}, aLIGO~\cite{TheLIGOScientific:2014jea} and SKA~\cite{Carilli:2004nx}. The corresponding constraints from the upper bound of stochastic GW backgrounds can be found in Ref.~\cite{NANOGrav:2021flc} ~(NANOGrav), Ref.~\cite{Xue:2021gyq}~(PPTA) and Ref.~\cite{Romero:2021kby}~(LIGO-Virgo). In this $Letter$, we propose a novel method to constrain the PT parameters with curvature perturations, and obtain strict constraints on the PTs below $100\,$GeV.
	
	It is well-known that quantum fluctutions during inflation successfully generate primordial perturbations required to explain the cosmic microwave background~(CMB) temperature anisotropies and the large-scale structure~\cite{Lewis:1999bs,Bernardeau:2001qr}. We find that the randomness of the quantum tunneling process during the PTs can also induce curvature perturbations both outside and inside the Hubble horizon, which can be orders of magnitude larger than the primordial perturbations for strong and slow PTs~\cite{Liu:2021svg}. The asynchronism
	of vacuum decay leads to some difference in the probability of the false vacuum and then affects the averaged equation of state within different Hubble horizons during the PTs, finally induces the curvature perturbations at superhorizon scales after the PTs. This kind of curvature perturbations is produced after inflation and independent of that produced during inflation. 
	Since in general the PTs happen in the early Universe when the comoving Hubble horizon is much smaller than the CMB scales, the perturbations induced by the PTs peak at small scales and do not affect the CMB observables for the scales of $10^{-4}\,\mathrm{Mpc}^{-1}<k<1\,\mathrm{Mpc}^{-1}$~\cite{Planck:2018vyg}. 
	At small scales, the current limits of the cosmic microwave background~(CMB) spectral distortions, the helium abundance, and the ultra-compact minihalo~(UCMH) abundance give constraints on curvature perturbations for $1\,\mathrm{Mpc}^{-1}<k<10^{7}\,\mathrm{Mpc}^{-1}$, which allow us to constrain the PTs that happen at higher energy scales.
	In Refs.~\cite{Emami:2017fiy,Gow:2020bzo}, these upper bounds on $\mathcal{P}_{\mathcal{R}}$ are employed to constrain the inflationary models and the primordial black hole abundance, and we here for the first time utilize these upper bounds to constrain the PT parameters, which largely extends the current constraint from the mostly discussed GWs from bubble collisions and sound waves.  
	
	For convenience, we choose $c=8\pi G=1$ throughout this letter.
	
	\emph{Curvature perturbations induced from the PTs}.
	\begin{figure*}
		\flushleft
		$\qquad$
		\includegraphics[height=1.9in]{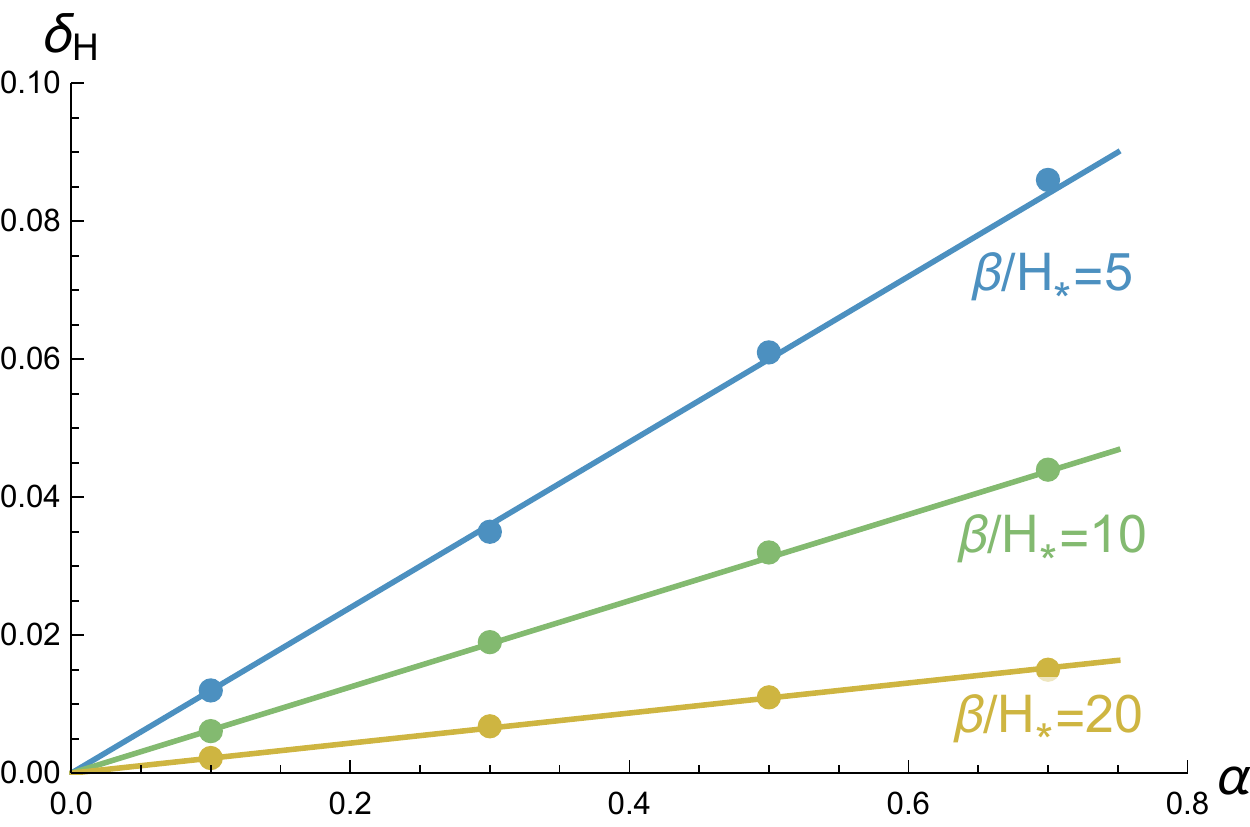}
		$\qquad$
		\includegraphics[height=1.9in]{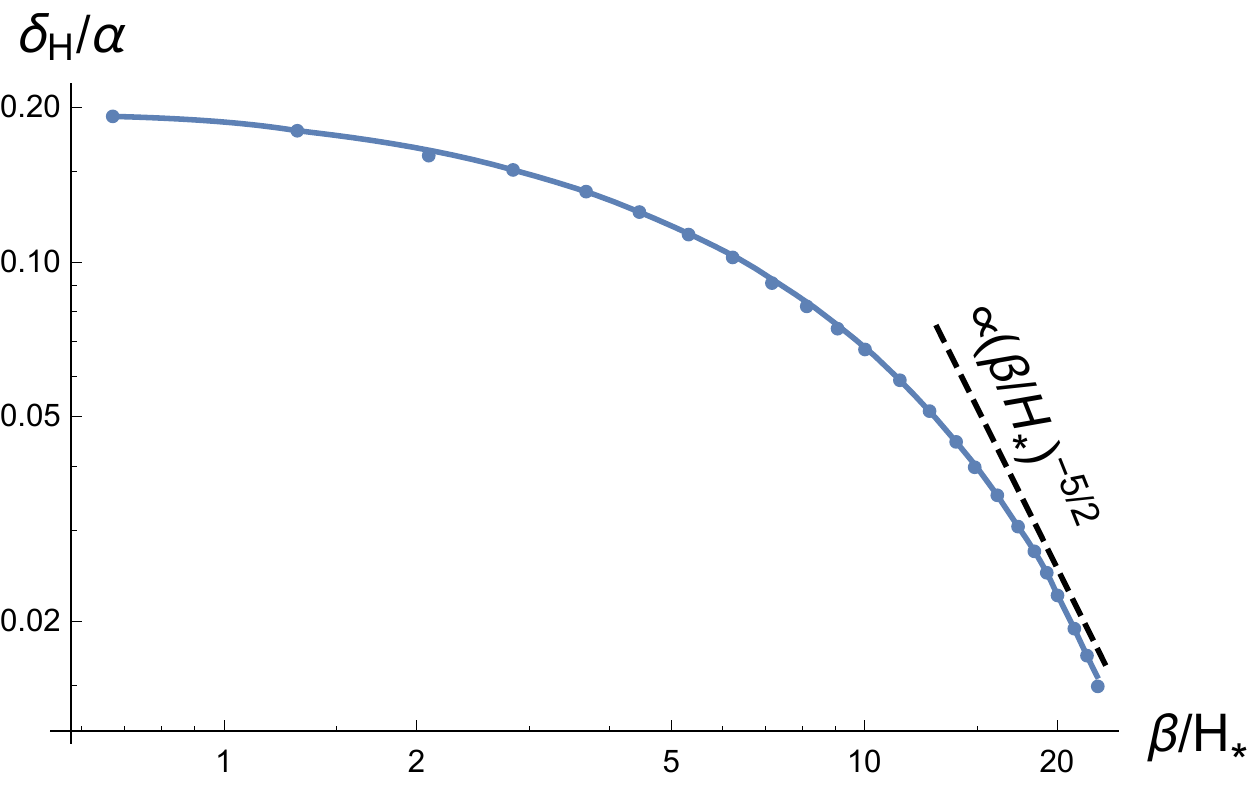}
		\caption{Left: The linear dependence of $\delta_{H}$ on $\alpha$ for $\beta/H_{*}=5$, $10$ and $20$, respectively. Right: The result of $f(\beta/H_{*})\equiv \delta_{H}/\alpha$ from numerical methods, which asymptotically approaches the analytical relation $\delta_{H}/\alpha\propto (\beta/H_{*})^{-5/2}$.}
		\label{fig:alphafbeta}
	\end{figure*}
	The nucleation rate of true vacuum bubbles generally takes the exponential form~\cite{Coleman:1977py,Enqvist:1991xw}
	\begin{equation}
		\label{eq:Gamma}
		\Gamma(t)=\Gamma_{0}e^{\beta t}\,,
	\end{equation}
	where $\Gamma_{0}$ and $\beta$ are approximately constants. For $\beta\gg 1$, $\beta^{-1}$ is also an estimation of the PT duration time. The averaged probability of the false vacuum, $F(t)$, reads~\cite{Turner:1992tz}
	\begin{equation}
		\label{eq:Ftn}
		F(t)=\exp \left[-\frac{4 \pi}{3} \int_{t_{i}}^{t} \mathrm{~d} t^{\prime} \Gamma\left(t^{\prime}\right) a^{3}\left(t^{\prime}\right) r^{3}\left(t, t^{\prime}\right)\right]\,,
	\end{equation}
	where $t_{i}$ is the time when quantum tunneling starts and $r(t,t')\equiv\int_{t'}^{t} a^{-1}(\tau) d\tau$ is the comoving radius of true vacuum bubbles. Before $t_{i}$ the field settles in the false vacuum so that $F(t<t_{i})=1$. After $t_{i}$, $F(t)$ decreases and the vacuum energy transfers into bubble walls and background plasma. 
	The PT temperature $T_{*}$ is evaluated at the percolation time $t_{\mathrm{p}}$ with $F(t_{\mathrm{p}})=0.7$. The Friedmann equation and the equations of motion read
	\begin{equation}\label{eq:fried}
		H^{2}=\frac{1}{3}(\rho_{\mathrm{r}}+\rho_{\mathrm{w}}+\rho_{\mathrm{v}})\,,
	\end{equation}
	\begin{equation}\label{eq:false}
		\rho_{\mathrm{v}}=F(t)\Delta V\,,
	\end{equation}
	\begin{equation}\label{eq:rho}
		\frac{d(\rho_{\mathrm{r}}+\rho_{\mathrm{w}})}{dt}+4H(\rho_{\mathrm{r}}+\rho_{\mathrm{w}})=\left(-\frac{d\rho_{\mathrm{v}}}{dt}\right)\,,
	\end{equation}
	where $H$ is the Hubble parameter, $\rho_{\mathrm{r}}$, $\rho_{\mathrm{w}}$ and $\rho_{\mathrm{v}}$ are the energy densities of background radiation, bubble walls and the false vacuum, respectively. $\Delta V$ is the energy density difference between the false and true vacua, and the true vacuum has been normalized to be the zero-point vacuum energy. In Eq.~\eqref{eq:rho} we assume the bubble wall velocity is close to $1$ so that bubble walls and the background plasma are both ultra-relativistic matter. The left-hand side of Eq.~\eqref{eq:rho} represents the evolution of ultra-relativistic matter in the expanding Universe and the right-hand side results from the decay of the vacuum energy. 
	
	Since in the expanding Universe $\rho_{\mathrm{r}}$ and $\rho_{\mathrm{w}}$ decrease as $a^{-4}$ while $\Delta V$ remains almost constant, so if the false vacuum decays later, the energy density becomes larger after a PT. Thus, the asynchronism of the vacuum decay process in different Hubble horizons induces superhorizon curvature perturbations. To analytically estimate perturbations of the total energy density $\delta\rho/\rho$, we notice that  the uncertainty of the vacuum decay time is about $\beta^{-1}$ in each region of the volume $\beta^{-3}$, then equations of motion~(\ref{eq:fried},\ref{eq:false},\ref{eq:rho}) imply the corresponding amplitude of $\delta\rho/\rho$ is proportional to $\alpha\beta^{-1}$ at the length scale of $\beta^{-1}$.
	Since the vacuum decay process becomes irrelevant at a length scale larger than $\beta^{-1}$, then the causality requires $\delta\rho/\rho\propto k^{3/2}$ at the infrared region.
	Then, the estimation of the variance of $\delta\rho/\rho$ at the Hubble horizon scale can be described as, 
	\begin{equation}\label{eq:5/2}
		\delta_{H}\equiv\sqrt{\langle(\delta\rho/\rho)^{2}\rangle_{H}}\propto \alpha(\beta/H_{*})^{-5/2}\,,
	\end{equation}
	where the subscript $H$ denotes the average over a Hubble horizon, $\beta/H_{*}\equiv \beta/H(t_{\mathrm{p}})$ and $\alpha\equiv\Delta V/\rho_{\mathrm{r}}(t_{\mathrm{p}})$ are two important universal parameters of the PTs, representing the speed and strength of the PTs, respectively. Note that the approximation~\eqref{eq:5/2} is valid in the case $\alpha<1$ and $\beta/H_{*}\gg 1$.
	
	The approximation~\eqref{eq:5/2} is then verified by numerical methods.
	We apply the postponed vacuum decay mechanism proposed in our previous work~\cite{Liu:2021svg} to numerically estimate $\delta\rho/\rho$. Let $P(t_{n})$ denote the probability that no vacuum bubbles nucleate before $t_{n}$ in a Hubble horizon, and the expression of $P(t_{n})$ is obtained from Eq.~\eqref{eq:Gamma},
	\begin{equation}
		\label{eq:prob}
		P(t_{n})=\exp\left[-\frac{4\pi}{3}\int_{t_{i}}^{t_{n}}\frac{a^{3}(t)}{a^{3}(t_{e})}H^{-3}(t_{e})\Gamma(t) dt\right]\,,
	\end{equation}
	where $t_{e}$ is the time when the false vacuum completely decays inside that Hubble-sized region.
	
	\begin{figure*}
		\flushleft\includegraphics[width=2.7in]{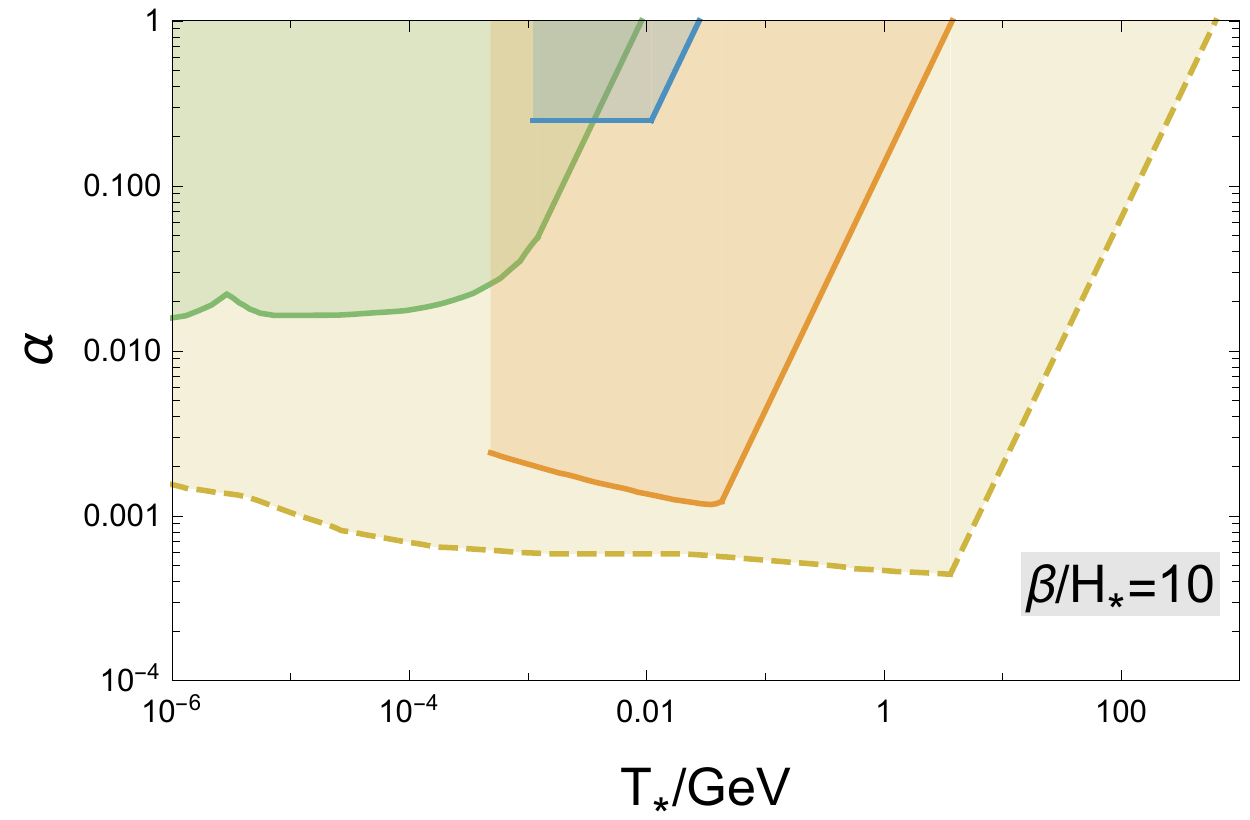}
		$\quad$
		\includegraphics[width=2.7in]{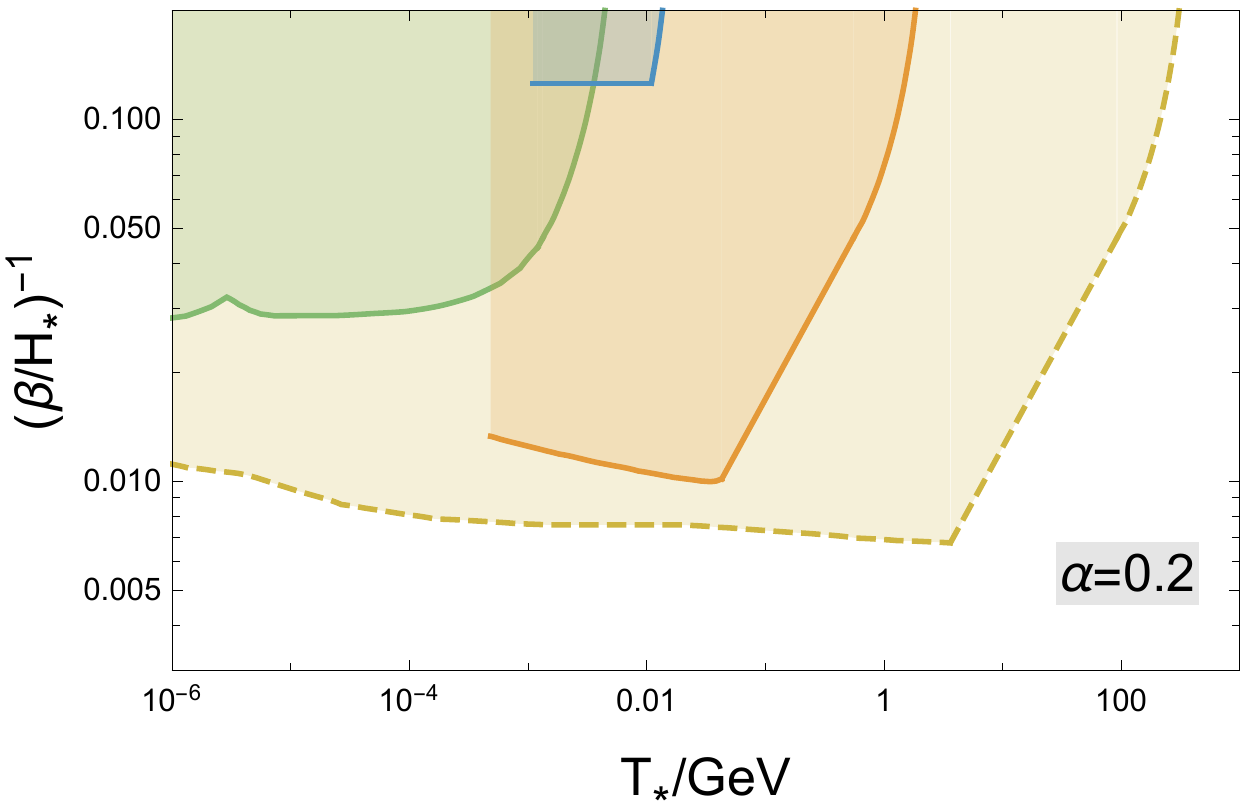}
		\raisebox{0.075\linewidth}{\includegraphics[width=1.3in]{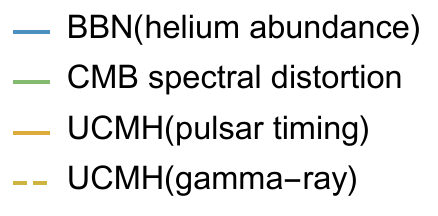}}
		\caption{The left and right panel  show the constraint on $\alpha$ and $\beta/H_{*}$ in the range $T_{*}\sim 10^{-6}-10^{3}\,$GeV, respectively, where we set $\beta/H_{*}=10$~(left) and $\alpha=0.2$~(right).}
		\label{fig:abT}
	\end{figure*}
	
	The delay of vacuum decay results in an increase of $\rho$ after the PTs, while the probability $P(t_{n})$ decreases for larger delay time $t_{n}$. We apply $P(t_{n})=0.16$ to estimate the variance of $\delta\rho/\rho$~(The constant $0.16$ comes from the probability of $x>\sigma$, where $x$ is a random variable satisfying the Gaussian distribution $X\sim \mathcal{N}(0,\sigma^{2})$. Even if the distribution of $\delta\rho/\rho$ may deviate from a Gaussian distribution, we can still use $P=0.16$ to estimate the variance of $\delta\rho/\rho$).  We solve Eqs.~(\ref{eq:fried},\ref{eq:false},\ref{eq:rho}) self-consistently and numerically obtain $\delta\rho/\rho$. In Fig.~\ref{fig:alphafbeta}, the left panel shows $\delta_{H}$ is proportional to $\alpha$ for different $\beta/H_{*}$, and the right panel shows the numerical result of $f(\beta/H_{*})\equiv\delta_{H}/\alpha$, where $f(\beta/H_{*})$ approaches to the asymptotic relation $(\beta/H_{*})^{-5/2}$. Thus, the analytical estimation~\eqref{eq:5/2} is verified numerically.
	
	The power spectrum of curvature perturbations, $\mathcal{P}_{\mathcal{R}}(k)$, is directly related to $\delta_{H}$ by~\cite{Carr:2009jm}
	\begin{equation}\label{eq:deltaH}
		\delta^{2}_{H}=\frac{16}{81}\int_{0}^{\infty}\frac{dk}{k}\left(kR_{H}\right)^{4}W^{2}\left(k,R_{H}\right)\,\mathcal{P}_{\mathcal{R}}(k)\,,
	\end{equation} 
	where $R_{H}=1/(aH)$ is the comoving Hubble radius at the end of the PTs, we apply a Gaussian form window function $W(k,R_{H})=\exp(-k^{2}R_{H}^{2})$.
	Since causality requires $\mathcal{P}_{\mathcal{R}}(k)\propto k^{3}$ for $k\ll \beta^{-1}$, we can finally obtain the approximate result of $\mathcal{P}_{\mathcal{R}}(k)$ in terms of the the numerical results of $f(\beta/H_{*})$ shown in Fig.~\ref{fig:alphafbeta},  
	\begin{equation}\label{eq:PRf}
		\mathcal{P}_{\mathcal{R}}(k)=34.5\alpha^{2}\left(f(\beta/H_{*})\right)^{2}(kR_{H})^{3}\,,
	\end{equation} 
	where the constant $34.5$ is obtained from Eq.~\eqref{eq:deltaH} and the numerical results in Fig.~\ref{fig:alphafbeta}. Since $\mathcal{P}_{\mathcal{R}}(k)\propto k^{3}$, at the CMB scales, curvature perturbations induced during the PTs are negligible compared to the primordial perturbations from inflation. 
	
	In the above derivations, we simply assume that at small scales, the primordial perturbations from inflation are negligible compared to that induced by the PTs so that Eq.~\eqref{eq:PRf} only contains the information of the PTs. This is the most natural case since at the CMB scales $\mathcal{P}_{\mathcal{R}}(k)\approx 2\times10^{-9}$ is nearly scale-invariant, which is also predicted by most inflationary models~\cite{Lyth:1998xn}. Therefore, the PTs alone can induce large curvature perturbations even if the curvature perturbations before the PTs are close to zero. However, it is possible that the small-scale $\mathcal{P}_{\mathcal{R}}$ is already very large before the PTs, as numerically studied in Ref.~\cite{Jinno:2021ury}, false vacuum decays later in high-temperature regions so that the  induced curvature perturbations should become larger than the prediction of Eq.~\eqref{eq:PRf}.
	
	\emph{Constraints on the PT parameters}.
	Since the PTs are expected to induce large curvature perturbations at small scales, the upper bounds on $\mathcal{P}_{\mathcal{R}}(k)$ can be converted into constraints on the PT parameters $\alpha$, $\beta/H_{*}$ and $T_{*}$. The observation of CMB and large-scale structure give strict constraints on the power spectrum of curvature perturbations, $\mathcal{P}_{\mathcal{R}}(k)\sim 2\times 10^{-9}$ at large scales  $k\lesssim \mathcal{O}(1)\,\mathrm{Mpc}^{-1}$. At smaller scales, $\mathcal{P}_{\mathcal{R}}(k)$ is constrained by other observables, such as\\
	$1.$ the limits of the CMB spectral distortions, \\
	$2.$ the helium abundance, \\
	$3.$ the pulsar timing constraint on the UCMH abundance, \\
	$4.$ the gamma-ray constraint on the UCMH abundance.
	
	Setting the scale factor at present, $a_{0}=1$, then in the radiation-dominated Universe, $k/(10^{4}\,\mathrm{Mpc}^{-1})\sim T_{*}/(1.1\,\mathrm{MeV})$, which roughly determines the range of $T_{*}$ constrained by each upper bounds of $\mathcal{P}_{\mathcal{R}}(k)$.
	We set the cutoff of Eq.~\eqref{eq:PRf} at the Hubble horizon scale $k=R_{H}^{-1}$ to obtain a conservative
	 estimation.
	In Fig.~\ref{fig:abT} we show the constraints on $\alpha$ and $\beta/H_{*}$ in the range $T_{*}\sim 10^{-6}-10^{3}\,$GeV from each constraint of $\mathcal{P}_{\mathcal{R}}(k)$.
	The limit of CMB spectral distortions, including $y$-distortion and $\mu$-distortion, implies $\mathcal{P}_{\mathcal{R}}(k)\lesssim 10^{-4}$ for the scales of $1\,\mathrm{Mpc}^{-1}\lesssim k\lesssim 10^{4}\,\mathrm{Mpc}^{-1}$~\cite{Chluba:2012we,Chluba:2015bqa,Lucca:2019rxf}, which gives constraints on the PTs with $T_{*}<1\,$MeV, as shown in the green lines. In the range $10^{4}\,\mathrm{Mpc}^{-1}\lesssim k\lesssim 10^{5}\,\mathrm{Mpc}^{-1}$, $\mathcal{P}_{\mathcal{R}}(k)$ has to be smaller than $0.01$ to avoid violating the BBN process and the prediction of primordial helium abundance~\cite{Jeong:2014gna,Nakama:2014vla,Inomata:2016uip}. The blue lines show the constraint from the observed helium abundance, constraining the PTs with $T_{*}\sim 1\,\mathrm{MeV}-10\,$MeV. The orange lines show the constraint from the limit of UCMH abundance, constraining the PTs with $T_{*}\sim 0.3\,\mathrm{MeV}-1\,$GeV. The authors of Refs.~\cite{Clark:2015sha,Clark:2015tha} find that UCMHs produce an observable period derivative of pulsars, and they apply the pulsar timing data to give the upper bound $\mathcal{P}_{\mathcal{R}}(k)\lesssim 10^{-6}$ for the scales of $4\times10^{3}\,\mathrm{Mpc}^{-1}\lesssim k\lesssim 4\times10^{5}\,\mathrm{Mpc}^{-1}$ with the redshift of the UCMH collapse $z_{c}=1000$ and detection threshold $s = 10\,ns$.
	The constraint presented in the dashed lines is from the non-observation of gamma-ray in UCMHs by $Fermi$ Large Area Telescope, which is valid in the case that WIMPs explain the nature of dark matter~\cite{Bringmann:2011ut,Nakama:2017qac,Kawasaki:2021yek,Delos:2018ueo}. Here, we apply the mass of WIMP particles $m_{\chi}=1\,$TeV, the annihilation cross section $\langle\sigma v\rangle=3\times 10^{-26}\,\mathrm{cm}^{3}s^{-1}$ and the UCMH redshift $z_{c}=1000$ as in Fig.6 of Ref.~\cite{Bringmann:2011ut}. This upper bound imposes a very strong constraint on all PTs below $100\,$GeV. All the constraints above are evaluated at the $95\%$ confidence level. 
	
	\begin{figure}
		\flushleft\includegraphics[width=3.2in]{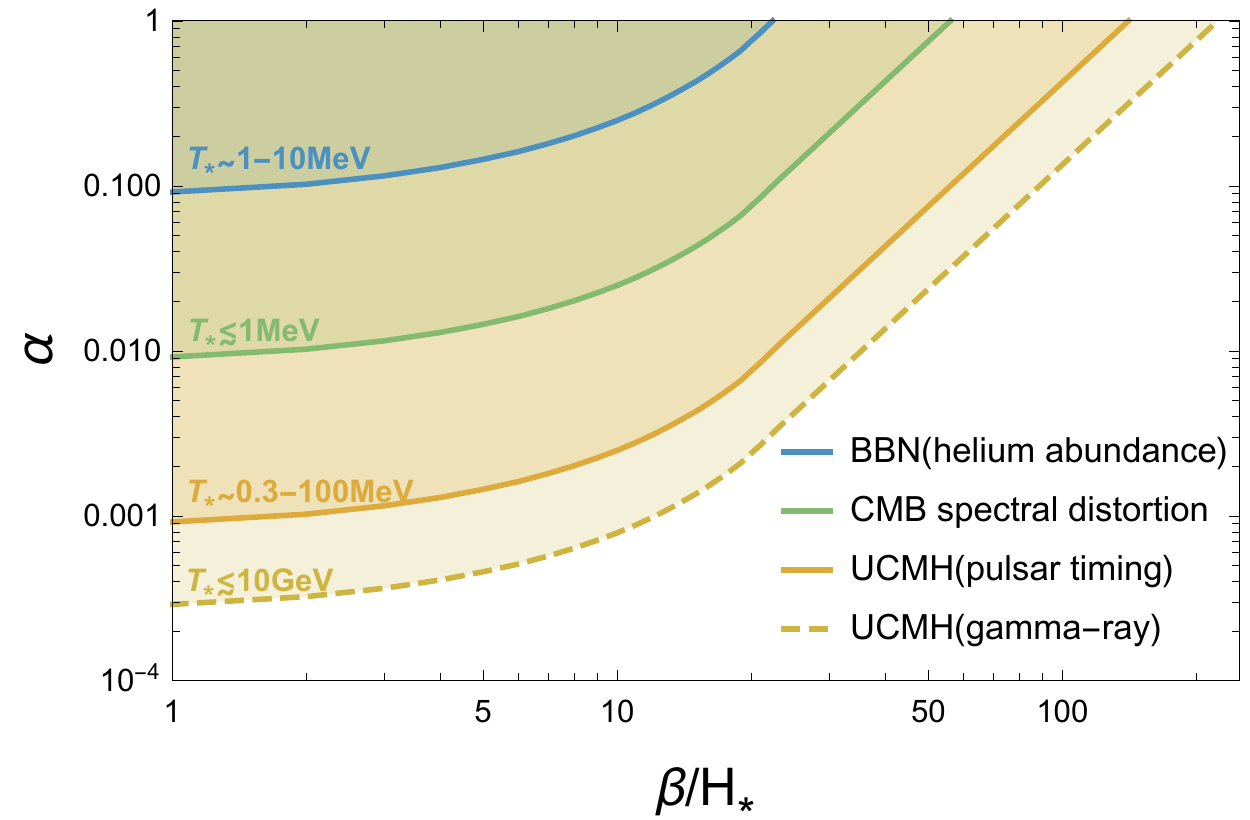}
		\caption{The excluded parameter spaces of $\alpha$ and $\beta/H_{*}$ obtained from each constraint on $\mathcal{P}_{\mathcal{R}}(k)$. }
		\label{fig:abcons}
	\end{figure}
	
	In Fig.~\ref{fig:abcons}, we show the excluded parameter spaces in the $\alpha-\beta/H_{*}$ plane and the corresponding application ranges of $T_{*}$ for each upper bound on $\mathcal{P}_{\mathcal{R}}(k)$. We obtain strict constraints mainly on the QCD first-order PTs~\cite{Schwarz:2009ii,Gao:2021nwz}\footnote{ The standard QCD phase transition is cross-over, but it becomes first-order in case of sufficiently large neutrino chemical potential~\cite{Schwarz:2009ii}.} and the low-scale dark PTs~\cite{Breitbach:2018ddu}. The constraint in this work is more strict on $\alpha$ than $\beta/H_{*}$, as Eq.~\ref{eq:PRf} implies. The most strict constraint on $\alpha$ is given by the limit of the UCMH abundance. In the case of  $\beta/H_{*}\lesssim 10$, the parameter space $\alpha\gtrsim 2\times 10^{-3}$~($\alpha\gtrsim6\times 10^{-4}$) has been excluded model-independently~(model-dependently) for $0.3\,\mathrm{MeV}\lesssim T_{*}\lesssim 100\, \mathrm{MeV}$~($T_{*}\lesssim 10\, \mathrm{GeV}$). Depending on the nature of dark matter, we also give the constraint on electroweak first-order PTs. In the case of $T_{*}=100\,$GeV, we obtain $\alpha\lesssim 0.1$ for $\beta/H_{*}=10$ and $\beta/H_{*}\gtrsim 20$ for $\alpha=0.2$, as shown in Fig.~\ref{fig:abT}. As a comparison, the current constraint from the PTA experiments gives $\alpha\lesssim 0.1$ for $\beta/H_{*}=10$ in the range $0.1\,\mathrm{MeV}\lesssim T_{*}\lesssim 0.1\,\mathrm{GeV}$ at the $95\%$ confidence level~\cite{Xue:2021gyq,NANOGrav:2021flc}. Ref.~\cite{Bai:2021ibt} obtains $\alpha\lesssim \mathcal{O}(0.01)$ for the PTs after the BBN process with $T_{*}\lesssim 1\,\mathrm{MeV}$, where the authors consider the decay product of false vacuum affects the effective number of relativistic degrees of freedom and the Helium abundance.
	Compared to the previous constraints, the constraint given in this work is much more strict on $\alpha$ and the application range of $T_{*}$ is also extended.
	
	\emph{Conclusion and discussion}.
	We quantitatively investigate the superhorizon curvature perturbations induced by the first-order PTs, and for the first time give constraints on the PT parameters using various upper bounds of $\mathcal{P}_{\mathcal{R}}(k)$. The  asynchronism of vacuum decay in different Hubble horizons induces large superhorizon curvature perturbations, which can be constrained by various observational results, such as the limits of the CMB spectral distortions, the helium abundance and the UCMH abundance. This work gives strict constraints on $\alpha$ and $\beta/H_{*}$ for the PT temperature below $100$GeV, including dark PTs, QCD first-order PTs and electroweak first-order PTs. The result largely expands the currently excluded parameter spaces obtained from the non-observation of the stochastic GW backgrounds, and give strict constraints especially on the low-temperature PTs and the weak PTs.
	
	In the near future, the space-based  GW detectors, such as LISA and $Taiji$, are sensitive to the PTs around the electroweak energy scale.  If the future GW observables may conflict with the model-dependent constraint~(the dashed line shown in Fig.~\ref{fig:abT} and Fig.~\ref{fig:abcons}), the discovery of GWs from the PTs in turn gives constraints on the annihilation cross section of the dark matter models.
	
	
	To avoid the overproduction of primordial black holes, curvature perturbations are loosely constrained as $\mathcal{P}_{\mathcal{R}}(k)<\mathcal{O}(0.01)$ over a wide range of scales $10^{-2}\,\mathrm{Mpc}^{-1}\lesssim k\lesssim10^{16}\,\mathrm{Mpc}^{-1}$. 
	Since the corresponding excluded parameter space in this case is rather small, we qualitatively conclude here that the strong PTs with the duration time comparable to $H^{-1}$ do not happen below $T_{*}<10^{9}\,$GeV.

	The semi-analytical result shown in Eq.~\ref{eq:PRf} is a conservative estimation since we only apply the averaged false vacuum probability in the overdense regions after $t_{n}$. Considering the accurate dynamics, including false vacuum probability and vector perturations, may result in larger curvature perturbations. The accurate result requires numerical simulations in the expanding Universe and the volume of the lattice should be much larger than $\beta^{-3}$.  We leave this topic and the corresponding constraints for future work.
	
	\emph{Acknowledgments}
	This work is supported in part by the National Key Research and Development Program of China Grants No. 2020YFC2201501, No. 2021YFC2203002 and 2021YFC2203004, in part by the National Natural Science Foundation of China Grants No. 11851302, No. 11947302, No. 11991052, No. 11821505, No. 12105060, No. 12105344 and No. 12075297, in part by the Science Research Grants from the China Manned Space Project with NO. CMS-CSST-2021-B01,
	in part by the Key Research Program of the CAS Grant No. XDPB15 and by Key Research Program of Frontier Sciences, CAS. Ligong Bian is supported by the National Natural Science Foundation of China under the grants Nos. 12075041, 12047564, 12147102, and the Fundamental Research Funds for the Central Universities of China (No. 2021CDJQY-011 and No. 2020CDJQY-Z003),  and Chongqing Natural Science Foundation (Grants No.cstc2020jcyj-msxmX0814).
	\bibliography{PRPT}
\end{document}